\documentclass[10pt]{article}
\usepackage{amsfonts,amssymb,latexsym,amsmath,amscd}
\usepackage{times,mathptm}
\usepackage[ps2pdf,colorlinks]{hyperref}
\usepackage{anysize}
\marginsize{1in}{1in}{1in}{1in}

\newcommand{\ket}[1]{|#1\rangle}
\newcommand{\bra}[1]{\langle#1|}

\begin{document}

\title{Note on the Khaneja Glaser Decomposition}

\author{Stephen S. Bullock\footnote{Mathematical and Computational Sciences 
Division, National Institute of Standards and 
Technology, Gaithersburg, MD,  20899, {\tt stephen.bullock@nist.gov}}
}

\maketitle

\begin{abstract}
Recently, Vatan and Williams utilize a matrix decomposition of
$SU(2^n)$ introduced by Khaneja and Glaser to produce {\tt CNOT}-efficient
circuits for arbitrary three-qubit unitary evolutions.  In this note,
we place the Khaneja Glaser Decomposition ({\tt KGD}) in context as a
$SU(2^n)=KAK$ decomposition by proving that its Cartan involution
is type {\bf AIII}, given $n \geq 3$.  The standard 
type {\bf AIII} involution produces the
Cosine-Sine Decomposition (CSD), a well-known decomposition
in numerical linear algebra which may be computed
using mature, stable algorithms.  In the course of our
proof that the new decomposition is type {\bf AIII}, we further establish
the following.  Khaneja and Glaser allow for a particular degree of freedom,
namely the choice of a commutative algebra $\mathfrak{a}$, in their
construction.  Let $\chi_1^n$ be a {\tt SWAP} gate applied on qubits
$1$, $n$.  Then $\chi_1^n v \chi_1^n=k_1\; a \; k_2$ is a KGD
for $\mathfrak{a}=\mbox{span}_{\mathbb{R}}
\{ \chi_1^n ( \ket{j}\bra{N-j-1} -\ket{N-j-1}\bra{j}) \chi_1^n \}$
if and only if $v=(\chi_1^n k_1 \chi_1^n)
(\chi_1^n a \chi_1^n)(\chi_1^n k_2 \chi_1^n)$ is a CSD.
\end{abstract}

Any fixed-time closed-system evolution of the $n$-qubit state space may
be modelled mathematically by multiplication of a state vector $\ket{\psi}$
by some $N \times N$ unitary matrix $v$, where $N=2^n$ throughout.
By choice of global phase, we may multiply $v$ by $\mbox{det}(v)^{1/N}$,
so that without loss of generality $v \in SU(N)$, the Lie group 
\cite{Knapp:98,Helgason:01} of determinant 
one unitary matrices.  Matrix decompositions
are algorithms for factoring matrices.  In the context of qubit dynamics,
such a decomposition would split an evolution $v$ into component
subevolutions.

A matrix decomposition may be proven by explicitly specifying an algorithm
that computes it.  Alternately, several theorems in Lie theory posit
factorizations of a group across certain subgroups.  These may in essence
be viewed as meta-decomposition theorems; they often allow many degrees of
freedom in the choice of subgroups and the group being factored.
Without describing the appropriate hypotheses, 
we mention examples such as the global Cartan decomposition 
$G=\mbox{exp}(\mathfrak{p})K$,
the Iwasawa decomposition $G=NAK$, and its generalization the Langlands
decomposition $G=UMAK$.  For $G=Gl(n,\mathbb{C})$ the Lie group of all
invertible complex matrices, well-known algorithms often exist for the
outputs of these theorems.  For example, the global Cartan decomposition
outputs the usual polar decomposition which writes a matrix as a
product of a Hermitian and unitary matrix, while the Iwasawa decomposition
reduces to the $QR$ decomposition.

The $G=KAK$ metadecomposition theorem has seen several overt and
hidden applications in quantum computing.  A sample output is
$SU(2^1)=\{R_y(\theta)\} \{ R_z(\alpha)\} \{R_y(\theta)\}$, where
$R_y(\theta)=\cos \frac{\theta}{2} \ket{0}\bra{0}
+\sin \frac{\theta}{2} \ket{0}\bra{1}
-\sin \frac{\theta}{2} \ket{0}\bra{1}
+\cos \frac{\theta}{2} \ket{1}\bra{1}$ and
$R_z(\alpha)=
\mbox{e}^{-i \alpha/2}\ket{0}\bra{0}+\mbox{e}^{i\alpha/2}\ket{1}\bra{1}$.
This is the
factorization of any Bloch-sphere rotation into rotations about orthogonal
axes.  For compact groups such as $SU(N)$, the definitive statement of
the $G=KAK$ theorem is found in Helgason \cite[thm8.6,\S VII.8]{Helgason:01}.

The Khaneja Glaser decomposition (KGD) is constructed by an explicit
invocation of the theorem, with $G=SU(2^n)$.  In fact, one
formulation \cite[Cor.3]{KhanejaGlaser:01} requires two applications of
the theorem.  In this note, we discuss the statement of Corollary 2 ibid.
Also, we note that the Cosine Sine decomposition (CSD) 
\cite[pg.77]{GolubVL:96} \cite{PageWei:94}
of numerical linear algebra is 
the output of the theorem for one standard choice
of inputs for $G=SU(2^n)$.  Per the statement of the abstract,
these two matrix decompositions are closely related.  Indeed, one
results from the other after swapping the labels on the first and
last qubit.  We derive this result in context, using the $G=KAK$
language.  As such, this note is another instance of a Lie theoretic
decomposition specializing to matrix analysis.

The theorem has three inputs, each dependent on the last.  The first
is the Lie group $G$ with Lie algebra $\mathfrak{g}$.  Should
$G \subseteq Gl(n,\mathbb{C})$ as a closed subgroup, i.e. $G$ is linear,
then $\mathfrak{g}$ is the set (in fact vector space) of matrix logarithms of
elements of $G$.  The algebra operation in this case is given by
$[X,Y]=XY-YX$ for $X,Y \in \mathfrak{g}$.  The second input is a Cartan
involution on $\mathfrak{g}$.  In the case that $G$ is compact, we take
this to mean an $\mathbb{R}$-linear map $\theta: \mathfrak{g} \rightarrow
\mathfrak{g}$ is a Lie algebra homomoprhism ($\theta[X,Y]=[\theta X,
\theta Y]$) and an involution ($\theta^2 = I_{\mathfrak{g}}$.)  It is
now typical to write $\mathfrak{g}=\mathfrak{p} \oplus \mathfrak{k}$,
with $\mathfrak{p}$ the $-1$ eigenspace and $\mathfrak{k}$ the subalgebra
given by the $+1$ eigenspace.  The final input to the theorem is a
maximal commutative subalgebra $\mathfrak{a}$ of $\mathfrak{p}$.  We now
state the $G=KAK$ theorem for compact groups.  This statement is a consequence
of the citation and follows by ignoring the discussion of Weyl group
actions ibid.

\medskip

\noindent
{\bf Theorem:}
{\bf (Cf. \cite[thm8.6,\S VII.8]{Helgason:01})}
Let $U$ be a connected, compact Lie
group with semisimple Lie algebra 
$\mathfrak{u}$, let $\theta:\mathfrak{u} \rightarrow \mathfrak{u}$
a Cartan involution of $\mathfrak{u}$, and
let $\mathfrak{a} \subset \mathfrak{p}$ a commutative subalgebra which is
maximal with this property.  For the Lie group exponential, label
$K=\mbox{exp}(\mathfrak{k})$, $A=\mbox{exp}(\mathfrak{a})$.  Then
\begin{equation}
U = KAK = \{ \ k_1 \; a \; k_2 \; \; ; \; \; k_j \in K, j=1,2, a \in A \ \}
\end{equation}
Note that the factorization of any given $u \in U$ is not unique.

\medskip

As a brief motivation, we mention how the Singular Value Decomposition
(SVD) \cite[pg.70]{GolubVL:96} arises as an example of the above construction
for noncompact $G$ \cite[pg.397]{Knapp:98}.  
If $G=Gl(n,\mathbb{C})$, then the set of matrix
logarithms is $\mathfrak{g}=\mathbb{C}^{n \times n}$.  Take
$\theta(X)=-X^\dagger$, so that $\mathfrak{p}$ the $-1$ eigenspace coincides
with the set of Hermitian matrices and $\mathfrak{k}$ the $+1$ eigenspace
is the set of antiHermitian matrices.  Hence $K=\mbox{exp}(\mathfrak{k})$
is the unitary group, and the Cartan involution has formalized a polar
decomposition at the level of $\mathfrak{g}$.  Now a suitable choice
of $\mathfrak{a}$ would be the diagonal matrices, so that
$Gl(n,\mathbb{C})=KAK=U(n)AU(n)$ in this case would be the SVD,
up to ordering and positivity of the diagonal factor.  In fact, this
degree of freedom is accounted for by the Weyl group action in the
cited theorem.  We do not exploit it here.

The Cosine-Sine Decomposition (CSD) 
\cite[pg.77]{GolubVL:96} 
may also be viewed as an
example of this theorem.  All possible Cartan involutions
$\theta$ of semisimple real Lie algebras, both compact and noncompact, are
classified \cite[p.518]{Helgason:01}.  Such Cartan involutions,
very loosely analogous to a polar decompositions at the Lie algebra
level, are less rigid for compact
than noncompact algebras.  For on a noncompact algebra
(e.g. $\mathfrak{sl}(n,\mathbb{R})$,) all subalgebras fixed
by Cartan involutions are Lie algebra isomorphic 
(e.g. to $\mathfrak{so}(n)$).  In the compact case, this is
false.  Three types of $\mathfrak{k}$ algebras may arise for Cartan
involutions of $\mathfrak{su}(N)$.  The types are {\bf AI}, {\bf AII},
and {\bf AIII} respectively \cite[pg.518]{Helgason:01} corresponding to
$\mathfrak{k} \cong 
\mathfrak{so}(N)$, $\mathfrak{sp}(N/2)$, and $\mathfrak{s}[\mathfrak{u}(p)
\oplus \mathfrak{u}(q)], p+q=N$.  The CSD is a $SU(N)=KAK$ decomposition
whose Cartan involution is of type {\bf AIII}.

Indeed, take $I_{N/2,N/2}=\left( \begin{array}{rr} I_{N/2} & {\bf 0} \\
{\bf 0} & -I_{N/2} \\ \end{array} \right)
=(\sigma^z) \otimes I_{N/2}$.  Then the standard Cartan involution
in the compact case of type {\bf AIII} is
$\theta_{\bf AIII}(X)= I_{N/2,N/2} X I_{N/2,N/2}$ \cite[pg.452]{Helgason:01}.
Note that the $+1$ eigenspace $\mathfrak{k}_{\bf AIII}$ is
\begin{equation}
\mathfrak{k}_{\bf AIII}
=\mathfrak{s}[\mathfrak{u}(N/2) \oplus \mathfrak{u}(N/2)]
= \Bigg\{ \ \left( \begin{array}{rr} { X} & {\bf 0} \\
{\bf 0} & { Y} \\ \end{array} \right) \; ; \;
X,Y \in \mathfrak{u}(N), \mbox{tr}(X)+\mbox{tr}(Y)=0 \ \Bigg\}
\end{equation}
There is moreover listed a standard choice of maximal commutative subalgebra.
\begin{equation}
\mathfrak{a}_{\bf AIII} =
\Bigg\{ \ \left( \begin{array}{rr} {\bf 0} & -T \\
T & {\bf 0} \\ \end{array} \right) \; ; \; 
T=\mbox{diag}(t_0,t_1,\cdots,t_{N/2-1})
\ \Bigg\}
\end{equation}
Then the assertion that $SU(2^n)=K_{\bf AIII} A_{\bf AIII} K_{\bf AIII}$
is the CSD, where
$K_{\bf AIII}=\mbox{exp}(\mathfrak{k}_{\bf AIII})$ and
$A_{\bf AIII}=\mbox{exp}(\mathfrak{a}_{\bf AIII})$.  
For upon exponentiating, the metadecomposition theorem
for these inputs states that any $v \in SU(N)$ may be written
\begin{equation}
v = \left( \begin{array}{rr} u_1 & {\bf 0} \\ {\bf 0} & 
u_2 \\ \end{array} \right)
\left( \begin{array}{rr} \cos(T) & -\sin(T) \\ \sin(T) & 
\cos(T) \\ \end{array} \right)
\left( \begin{array}{rr} u_3 & {\bf 0} \\ {\bf 0} & u_4 \\ \end{array} \right)
\end{equation}
where $u_j$ is unitary for $j=1,2,3,4$,
the determinant of each factor is one, and $T$ is as in the definition of
$\mathfrak{a}_{\bf AIII}$.  Thus, if $\gamma$ is the namesake cosine-sine
matrix of the central factor, we recover the CSD by
$v=(u_1 \oplus u_2) \gamma (u_3 \oplus u_4)$.

\vbox{
For the remainder, we use the notations $A$, $K$, $\mathfrak{a}$, and
$\mathfrak{k}$ to correspond to the choices of metadecomposition input
made in the construction of the KGD
\cite{KhanejaGlaser:01}.  
Motivated by physical intuition in terms of spin chains, 
Khaneja and Glaser do not explicitly
formulate $\theta$ but rather specify the eigenspace $\mathfrak{k}$.
This is sufficient, since $\mathfrak{p}$
and hence $\theta$ may be recovered from $\mathfrak{k}$
as a Killing form \cite[pg.131]{Helgason:01} orthogonal complement.
The conventions for the KGD are then as follows:
\begin{itemize}
\item  $\mathfrak{k}=\mbox{span}_{\mathbb{R}}\{ X \otimes \sigma^z, 
Y \otimes I_2, i I_{N/2} \otimes \sigma^z \; ; \; 
X, Y \in \mathfrak{su}(N/2) \}$, per \cite[thm.3]{KhanejaGlaser:01}
\item  any suitable \cite[Cor.2]{KhanejaGlaser:01}
maximal commutative $\mathfrak{a} \subset \mathfrak{p}$
\end{itemize}
For the second item, we take $\chi_1^n$ to be {\tt SWAP}
on qubits $1$, $n$.  Label
\begin{equation}
\mathfrak{a}\ =\ \mbox{span}_{\mathbb{R}}
\big\{ \; \chi_1^n \; 
\big( \; \ket{j}\bra{N-j-1} -\ket{N-j-1}\bra{j} \; \big) \; \chi_1^n \ ; \
0 \leq j \leq N/2-1 \; \big\}
\end{equation}
This $\mathfrak{a}$ is not related to the specific
choice of commutative subalgebra of the typical
KGD \cite[Notation5]{KhanejaGlaser:01}.  Rather, it recovers
the CSD.
}

\medskip

\vbox{
\noindent
{\bf Proposition:}
Suppose $n\geq 3$ throughout.
\begin{enumerate}
\item \label{it:kalg}
$\mathfrak{k}_{\bf AIII} = \chi_1^n \; \mathfrak{k} \; \chi_1^n$
\item \label{it:Kgrp}
$S[U(N/2)\oplus U(N/2)]= \chi_1^n \; K \; \chi_1^n$
\item \label{it:aalg}
$\mathfrak{a}_{\bf AIII} = \chi_1^n \; \mathfrak{a} \; \chi_1^n$
\item \label{it:Agrp}
$A_{\bf AIII} = \chi_1^n\; A \; \chi_1^n$
\end{enumerate}
Thus for $v\in SU(N)$, we have $v=k_1 a k_2$ a Khaneja Glaser decomposition
if and only if 
$\chi_1^n k_j \chi_1^n = u_{2j-1} \oplus u_{2j} \in S[U(N/2) \oplus U(N/2)]$,
$j=1,2$ and $\chi_1^n a \chi_1^n = \gamma \in A_{\bf AIII}$, i.e.
if and only if
\begin{equation}
\chi_1^n v \chi_1^n \ = \ (\chi_1^n k_1 \chi_1^n)(\chi_1^n a \chi_1^n)
(\chi_1^n k_2 \chi_1^n)
\ = \ (u_1 \oplus u_2) \gamma (u_3 \oplus u_4)
\end{equation}
is a Cosine Sine decomposition.
}

\medskip

\noindent
{\bf Proof:}
For Items \ref{it:kalg} and \ref{it:aalg}, use the standard properties of 
a {\tt SWAP}.  For the remaining two items, recall that $SU(N)$ is a linear
algebraic group.  Hence the Lie exponential is given by the usual power
series for a matrix exponential, and generally for any $v \in SU(N)$ and
$X \in \mathfrak{su}(N)$ one has $\mbox{exp}(v X v^\dagger) =
v \; \mbox{exp}(X) \; v^\dagger$.  Now consider $X$ in each of the two
Lie algebras and $\chi_1^n = (\chi_1^n)^\dagger$.
\hspace*{\fill}{$\Box$}\smallskip

Although the commutative subalgebra $\mathfrak{a}$ in the argument above is
new, this is not an essential obstacle to recovering
the KGD from a numerical CSD.  For
the theory of Weyl group actions demands that all $\mathfrak{a}$
satisfying the hypothesis of the $G=KAK$ theorem are conjugate under
some $k \in K$ \cite{KhanejaGlaser:01,Helgason:01}.  
Thus, let $\tilde{\mathfrak{a}} = \chi_1^n \mathfrak{h}(n)
\chi_1^n$ be the Khaneja Glaser commutative subalgera $\mathfrak{h}(n)$
\cite[Notation5]{KhanejaGlaser:01}
up to qubit {\tt SWAP}.  Note that specifically
\begin{equation}
\tilde{\mathfrak{a}} \ = \ \mbox{span}_{\mathbb{R}}
\{ \; i \sigma^x \otimes (\sigma^x)^{b_2} \otimes (\sigma^x)^{b_3} \otimes
\cdots (\sigma^x)^{b_{n-3}} \otimes (\sigma^j)^{\otimes 2} \; ; \;
j=0,x,y,z, b_k \in \{0,1\}, 2 \leq k \leq n-3 \; \}
\end{equation}
Then the theory demands some 
$k \in S[U(N/2)\oplus U(N/2)]$ block diagonal so that
$k \tilde{\mathfrak{a}} k^\dagger = \mathfrak{a}_{\bf AIII}$.  
Indeed, an important technique in computing
matrix decompositions for two-qubit logic-circuit synthesis 
is transforming the commutative Lie algebra
$\mbox{span}_{\mathbb{R}} \{ i(\sigma^x)^{\otimes 2}, i(\sigma^y)^{\otimes 2},
i(\sigma^z)^{\otimes 2} \}$ into a diagonal Lie algebra
\cite{ZhangEtAl:03,BullockMarkov:03}.  This may
be accomplished for example by
\begin{equation}
E = \frac{1}{\sqrt{2}} \left( \begin{array}{rrrr}
1 & 0 & i & 0 \\ 0 & 1 & 0 & i \\ 0 & -1 & 0 & i \\ 1 & 0 & -i & 0 \\
\end{array} \right)
\end{equation}
Then directly $E^\dagger ( \sigma^x \otimes \sigma^x) E = \sigma^z \otimes
\sigma^z$, $E^\dagger (\sigma^y \otimes \sigma^y) E = 
-\sigma^z \otimes I_2$, and
$E^\dagger (\sigma^z \otimes \sigma^z)E=I_2 \otimes \sigma^z$.  
Moreover $H \sigma^z H = H \sigma^z H^\dagger = \sigma^x$,
for $H=\frac{1}{\sqrt{2}}\sum_{j,k=0}^1 (-1)^{jk}\ket{k}\bra{j}$ the
Hadamard map.  Also, if $S=\ket{0}\bra{0}+i \ket{1}\bra{1}$, then
$S (i \sigma^x) S^\dagger = i \sigma^y$.
Thus suppose
we take ${k}= S \otimes H^{\otimes (n-3)} \otimes E^\dagger$.  Then 
the matrix $k$ may be used to switch commutative subalgebras:
\begin{equation}
{k} \tilde{\mathfrak{a}} {k}^\dagger \ = \ 
(k \chi_1^n) \mathfrak{h}(n) (k \chi_1^n)^\dagger \ = \ 
\mbox{span}_{\mathbb{R}} \{ \;
(i \sigma^y) \otimes (\sigma^z)^{b_2} \otimes \cdots \otimes (\sigma^z)^{b_n}
\; ; \; b_2, b_3, \ldots, b_n = 0,1 \; \} \ = \ \mathfrak{a}_{\bf AIII}
\end{equation}
Thus this block-unitary $k \in S[U(N/2)\oplus U(N/2)]$ allows for
translation between the CSD and that KGD where the commutative subalgebra
is chosen as $\mathfrak{h}(n)$.

The CSD has been used for the design
of quantum logic circuits directly in 
\cite{Tucci:99,Mottonen:04,ShendeBullockMarkov:04}.
We refer to the latter for advice on obtaining a numerical implementation
of the CSD.
Thus, this document serves to translate earlier circuit constructions
\cite{VatanWilliams:04} in the KGD into more recent works.  Similar
translations might be possible from other applications of
the KGD, e.g. in control theory \cite{KhanejaGlaser:01}.

\end{document}